\documentclass[aps,prc,preprint,groupedaddress]{revtex4}
\usepackage{graphicx}
\def\Vec#1{\mbox{\boldmath $#1$}}

\begin{document}


\title{
Criterion for Bose-Einstein condensation in traps and self-bound systems
}


\author{T.~\textsc{Yamada}, Y.~\textsc{Funaki}$^1$, H.~\textsc{Horiuchi}$^{2,3}$, G.~\textsc{R\"opke}$^4$,
        P.~\textsc{Schuck}$^{5,6,7}$, and A.~\textsc{Tohsaki}$^2$}
\affiliation{Laboratory of Physics, Kanto Gakuin University, Yokohama 236-8501, Japan}
\affiliation{$^1$Nishina Center for Accelerator-Based Science, The Institute of Physical and Chemical Research (RIKEN), Wako 351-0098, Japan}
\affiliation{$^2$Research Center for Nuclear Physics (RCNP), Osaka University, Osaka 567-0047, Japan}
\affiliation{$^3$International Institute for Advanced Studies, Kizugawa 619-0225, Japan}
\affiliation{$^4$Institut f\"ur Physik, Universit\"at Rostock, D-18051 Rostock, Germany}
\affiliation{$^5$Institut de Physique Nucl\'eaire, CNRS, UMR 8608, Orsay, F-91406, France}
\affiliation{$^6$Universit\'e Paris-Sud, Orsay, F-91505, France} 
\affiliation{$^7$Laboratoire de Physique et Mod\'elisation des Milieux Condens\'es, CNRS et Universit\'e Joseph Fourier, 25 Av.~des Martyrs, BP 166, F-38042 Grenoble Cedex 9, France}


\date{\today}

\begin{abstract}
The internal one-particle density matrix is discussed for Bose-Einstein 
condensates
 with finite number of particles in a harmonic trap.
It is found that the outcome of the diagonalization of the single particle 
 density matrix depends on the choice of the
 internal coordinates:~The Pethick-Pitaevskii(PP)-type internal density matrix,
 whose analytical eigenvalues and eigenfunctions are evaluated,
 yields a fragmented condensate, while the Jacobi-type internal density matrix
 leads to an ideal condensate. 
We give a criterion for the choice of the internal coordinates:~In the 
macroscopic
 limit the internal density matrix should have the same eigenvalues and 
eigenfunctions
 as those of the corresponding ideal Bose-Einstein condensate in the 
laboratory frame, 
 this being a very physical condition.
One choice fulfilling this boundary condition is given by the internal Jacobi
 coordinates, while the internal
 coordinates with respect to the center of mass of PP do not satisfy this 
condition.
Based on our criterion, a general definition of the internal one-particle
 density matrix is presented in a self-bound system, consisting of interacting 
bosons.
\end{abstract}


\maketitle

One of the most amazing phenomena in quantum many-particle systems is Bose-Einstein condensation.
A characteristic feature of the phenomenon is the macroscopic occupation of a single-particle state.
The criterion of condensation in homogeneous interacting boson system 
 was given many years ago by Penrose and Onsanger~\cite{penrose56} and Yang~\cite{yang62}
 who introduced the concept of off-diagonal long-range order (ODLRO).
The condensate fraction is given by diagonalizing the one-particle density matrix
 which, in general, is defined in the laboratory frame.
A dominant eigenvalue of the one-particle density matrix implies ODLRO. 
In other words, a system shows Bose-Einstein condensation (BEC) if only one 
eigenvalue
 is of the order of the particle number $N$ in the system and the others
 are of the order $N^0=1$.
In the case of the existence of several large and comparable eigenvalues,
 the system is said to be a fragmented condensate~\cite{nozieres82,legget01}.
This conventional criterion works well, for example, for superconductivity of metals 
 at low temperature and superfluidity of liquid $^4$He, including recent realization
 of BEC of ultracold atoms in magnetic traps~\cite{dalfovo99}.

A controversy concerning the criterion has, however, arisen for the case
 of condensation with attractive interactions among bosonic particles.
Wilkin et al.~discussed the lowest excitation of a condensate with
 attractive interactions whose system rotates around its center of mass but keeps
 all relative degrees of freedom in its ground state~\cite{wilkin98}.
It was found that the corresponding one-particle density matrix in the laboratory frame
 has many eigenvalues of comparable size, and thus the system should be characterized
 as a fragmented condensate.
Pethick and Pitaevskii (PP) claimed that their result is incorrect because
 only the center-of-mass degree of freedom is excited and the internal
 degrees of freedom remain in the lowest states~\cite{pethick00}.
This is, indeed, physically reasonable.
They pointed out that a careful treatment is needed in formulating
 a criterion for Bose-Einstein condensation.
They proposed that a convenient way to describe correlations in the system
 is by defining an internal one-particle density matrix in which 
 the center-of-mass motion is eliminated, and
 that this definition should give a single eigenvalue of the order of 
 the number of particles.
Unfortunately, they presented only their idea and did not demonstrate 
explicit results
 of the diagonalization of their definition of the internal one-particle density matrix.
On the other hand, Gajda stressed that PP's criterion is not complete in
 the case of attractive systems whose center-of-mass motion has a large
 uncertainty, namely, whose center-of-mass extension significantly
 exceeds other length scales concerning the internal coordinates~\cite{gajda06}.

Recently, $\alpha$-particle 'condensation' in nuclear systems~\cite{tohsaki01} 
has attracted much interest.
A typical well established example is the $3\alpha$ condensate in the Hoyle state, 
 i.e.~the $0^+_2$ state in $^{12}$C, located just above the $3\alpha$ breakup threshold.
It is the finite size counterpart of macroscopic $\alpha$-particle condensation
 in infinite nuclear matter at low density~\cite{beyer00}.
The Hoyle state which was well described by several papers in
 the past~\cite{horiuchi74,fukushima78,uegaki77,descouvemont87}
 has, more recently, been investigated in more detail~\cite{tohsaki01,funaki03,chernykh07}
 and found to have about $1/3$ of saturation density of normal nuclei and  
 to be, in good approximation, a product state of three $\alpha$-particles,
 condensed into the lowest mean field $0S$-orbit with respect
 to the motion of their center-of-mass coordinates~\cite{matsumura04,yamada05}.
In addition current theoretical work indicates that also the $0^+_6$ state of $^{16}$O above
 the $4\alpha$ breakup threshold has dilute $4\alpha$ structure of condensate
 type~\cite{funaki08} and possibly states in heavier nuclei as well.

The character of these $\alpha$-particle condensates may be seen 
 as a few particle analogue to Bose-Einstein 
 condensation of ultracold atoms in magnetic traps~\cite{dalfovo99}.
However, some qualitatively different features exist between the two systems.
Besides the small number of particles, there is, for instance, the fact that
 the $\alpha$-particles form self-bound systems, where the center-of-mass motion
 can not be controlled by the external field, because the total wave function of the
 system should, in principle, be described in terms of only the internal coordinates,
 eliminating the center-of-mass degree of freedom.
Thus, the issue of the internal one-particle density matrix with a definite finite
 number of particles is very relevant in the study of $\alpha$-particle condensation
 in finite nuclear systems.   

The purpose of this paper is to give a suitable definition of the internal one-particle density
 matrix of a self-bound system with a finite particle number.
The criterion by PP seems to work well for this issue at first sight.
However, we will demonstrate that their criterion is not adequate, leading 
to a fragmented condensate, contrary to their initial objective.
For illustration, we treat a simple case, i.e.~the internal state of a Bose-Einstein
 condensate with finite particle number in a harmonic trap.
Let us
first consider the one-particle density matrix in the laboratory frame for an ideal
 Bose-Einstein condensate with $N$ spinless particles.
The result is trivial but instructive for studying the nature of the internal one-particle
 density matrix, as will be discussed later.  

The $N$-particle Hamiltonian in laboratory frame is presented as
\begin{eqnarray}
H = {\sum_{i=1}^{N} \frac{1}{2m}\Vec{p}_i^2} + \sum_{i=1}^N \frac{1}{2}m\omega^2 \Vec{r}_i^2.
\label{eq:total_hamiltonian}
\end{eqnarray}
The ground-state wave function of this system is expressed as a product of
 identical Gaussian single-particle wave functions, i.e.
\begin{eqnarray}
\Phi(\{\Vec{r}\}_{i=1}^N) &=& \prod_{i=1}^{N} \left( \frac{2\nu}{\pi} \right)^{3/4} \exp\left(-\nu\Vec{r}_i^2\right),
\label{eq:total_wf}
\end{eqnarray}
where $\nu=m\omega/2\hbar$, and $\{\Vec{r}\}_{i=1}^N$ denotes the set of the coordinates 
$\Vec{r}_i$ ($i=1\cdots,N$).
The one particle density matrix in the laboratory frame is defined as
\begin{eqnarray}
{\rho^{(1)}_{\rm Lab}(\Vec{r}, \Vec{r}')} = \int \prod_{i=2}^{N} d\Vec{r}_i 
       \Phi^*(\Vec{r},\{\Vec{r}_i\}_{i=2}^{N}) \Phi(\Vec{r}',\{\Vec{r}_i\}_{i=2}^{N}),\\
  = \left(\frac{2\nu}{\pi}\right)^{3/2} \exp\left[-\nu(\Vec{r}^2+\Vec{r}'^2)\right].\hspace*{14mm} 
\label{eq:one-body_density}
\end{eqnarray}
It is noted that the density matrix is independent of the number of particles $N$ and
 is separable with respect to $\Vec{r}$ and $\Vec{r}^{\prime}$.
The separability originates from the fact that the Hamiltonian is separable
 for $\Vec{r}_i$ and $\Vec{p}_i$ in Eq.~(\ref{eq:total_hamiltonian}).
 
The nature of the single particle orbits and their occupation probabilities in the relevant system
 can be obtained by solving the eigenvalue problem of the density matrix (\ref{eq:one-body_density}).
This is easily done, and we find that
 the density matrix has only one non zero eigenvalue with one eigenfunction, namely,
 the zero-node $S$-wave Gaussian $\varphi(\Vec{r})=(2\nu/\pi)^{3/4}\exp(-\nu\Vec{r}^2)$
 (or $0S$ harmonic oscillator wave function $R_{000}(\Vec{r},\nu)$)
 with 100~\% occupancy.
This means that all particles are condensed in that single orbit, i.e. an ideal Bose-Einstein
 condensation is realized in the laboratory frame. 
To say it again, this feature is independent of the number of particles $N$.

Next we consider the internal one-particle density matrix for the $N$-particle Bose-Einstein 
 condensation in a harmonic trap described by the wave function Eq.~(\ref{eq:total_wf}) 
 with the total Hamiltonian Eq.~(\ref{eq:total_hamiltonian}).
{\it Internal} means that the density is free from the center-of-mass
 coordinate of the system.
In the present paper, two kinds of internal coordinate sets are introduced,
 1)~coordinates with respect to the center of mass and 2)~Jacobi coordinates.
The former set was first considered by Pethick and Pitaevskii~\cite{pethick00} to define
 the internal one-particle density matrix.
We call it Pethick-Pitaevskii-type (PP-type) internal one-particle density matrix in the present paper.
For the latter set, we call it Jacobi-type density matrix.

\begin{center}
{\it PP-type internal one-particle density matrix}
\end{center}

In order to define an internal one-particle density matrix, Pethick and Pitaevskii adopted
 internal coordinates defined with respect to the center of mass~\cite{pethick00}.
The center-of-mass coordinate $\Vec{R}$ and the coordinate $\Vec{q}_i$
 of particle $i$ relative to the center of mass are given, respectively, by
\begin{eqnarray}
\Vec{R}=\frac{1}{N}\sum_{i=1}^N \Vec{r}_i, \hspace*{5mm}\Vec{q}_i=\Vec{r}_i-\Vec{R},\label{eq:coordinate_cm_rel}
\end{eqnarray}
where the definition of the center-of-mass coordinate $\Vec{R}$ implies that only $N-1$ of the
 $\Vec{q}_i$ are independent.
Here, we define the conjugate momenta $\Vec{\pi}_i$ and $\Vec{P}$
 for the coordinates $\Vec{q}_i$ and $\Vec{R}$, respectively.
Then, the total Hamiltonian in Eq.~(\ref{eq:total_hamiltonian}) is rewritten as
\begin{eqnarray}
&&H = H_{\rm int} + H_{\rm cm},\label{eq:PP_Hamiltonian}\\
&&H_{\rm int}=\frac{1}{2m} \left[ {\left(\frac{N-1}{N}\right) \sum_{i=1}^{N-1} \Vec{\pi}_{i}^2} - {\frac{2}{N} \sum_{i<i'=1}^{N-1} \Vec{\pi}_{i}\cdot\Vec{\pi}_{i'}} \right] \nonumber \\
&&\hspace*{15mm} + m\omega^2 \left[ \sum_{i=1}^{N-1}{\Vec{q}_i}^2 + \sum_{i<i'=1}^{N-1}\Vec{q}_i\cdot\Vec{q}_{i'}\right],\label{eq:PP_internal_Hamiltonian}\\
&&H_{\rm cm}=\frac{1}{2Nm}\Vec{P}^2 + \frac{1}{2}Nm\omega^2\Vec{R}^2,\label{eq:Hamiltonian_cm}
\end{eqnarray}
where $H_{\rm int}$ and $H_{\rm cm}$ denote the internal and center-of-mass Hamiltonians, respectively.

The total wave function in Eq.~(\ref{eq:total_wf}) is expressed as
\begin{eqnarray}
&&\Phi(\{\Vec{r}_i\}_{i=1}^{N}) = \frac{1}{N^{3/2}} \times  \Phi_{\rm int}(\{\Vec{q}_i\}_{i=1}^{N-1}) \Phi_{\rm cm}(\Vec{R}),\label{eq:wf_cm_rel}\\
&&\Phi_{\rm int}(\{\Vec{q}_i\}_{i=1}^{N-1})=\left( \frac{N\times (2\nu)^{N-1}}{\pi^{N-1}} \right)^{3/4} \nonumber \\
&& \hspace*{10mm} \times~\exp\left[-\sum_{i,i'=1}^{N-1} (\delta_{ii'}+1) \nu \Vec{q}_i \cdot \Vec{q}_{i'}\right],\label{eq:PP_wf_internal}\\
&&\Phi_{\rm cm}(\Vec{R})=\left( \frac{2N\nu}{\pi} \right)^{3/4} \exp(-N\nu\Vec{R}^2),\label{eq:PP_wf_cm}
\end{eqnarray}
where $\Phi_{\rm int}$ and $\Phi_{\rm cm}$ denote the internal and center-of-mass wave functions, respectively.

According to Pethick and Pitaevskii~\cite{pethick00}, the internal one-particle density matrix is given as~\cite{yamada08},
\begin{eqnarray}
\rho_{\rm int, PP}^{(1)}(\Vec{q},\Vec{q}^\prime) =\int {d\Vec{q}_2 \cdots d\Vec{q}_{N-1}} \hspace*{25mm} \nonumber \\
           \times~{\Phi^{*}_{\rm int}}(\Vec{q},\{\Vec{q}_i\}_{i=2}^{N-1}) {\Phi_{\rm int}}(\Vec{q}^\prime,\{\Vec{q}_i\}_{i=2}^{N-1}),\hspace*{5mm}\\
  = \left( \frac{N}{N-1} \right)^{3/2} \left( \frac{2\nu}{\pi} \right)^{3/2} \hspace*{35mm}\nonumber \\ 
     \times \exp\left[-\frac{3N-2}{2(N-1)}\nu(\Vec{q}^2+{\Vec{q}^\prime}^2) + \frac{N-2}{N-1}\nu\Vec{q}\cdot\Vec{q}^\prime \right].
\label{eq:one_body_den_int} 
\end{eqnarray}
It is noted that this density matrix depends on the number of particles $N$ and contains the
 cross term $\Vec{q}\cdot\Vec{q}^\prime$.
The origin of the cross term comes from the nonseparability of $\Vec{\pi}_i$ and $\Vec{q}_i$ in
 the internal Hamiltonian [see Eq.~(\ref{eq:PP_internal_Hamiltonian})].
One shall remark that the result of Eq.~(\ref{eq:one_body_den_int}) differs substantially from Eq.~(9)
 in the paper by Zinner and Jensen~\cite{zinner07}.

\begin{figure}
\begin{center}
\includegraphics*[scale=0.26,clip]{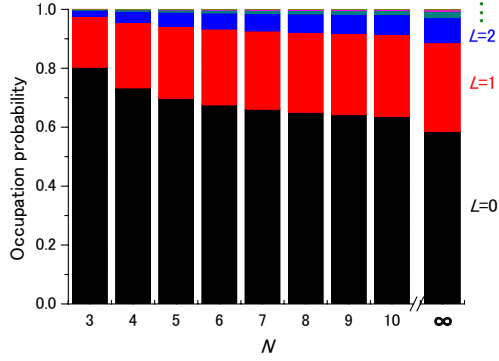}
\caption{
Spectra of the occupation probabilities $\Lambda^{(L)}_N$ 
 for the PP-type internal density matrix [see Eq.~(\ref{eq:PP_occupation_probablity_N_L})].
}
\label{fig:1}
\end{center}
\end{figure}

Let us discuss the nature of the internal one-particle density matrix 
$\rho_{\rm int}^{(1)}(\Vec{q},\Vec{q}^\prime)$.
First we study the single-particle orbits and their eigenvalues obtained by solving
 the eigenvalue problem,
\begin{eqnarray}
\int \rho_{\rm int,PP}^{(1)}(\Vec{q},\Vec{q}^\prime) \varphi(\Vec{q}) d\Vec{q}^\prime = \lambda \varphi(\Vec{q}). 
\end{eqnarray}
We find that this equation can be solved analytically.
The single-particle orbits $\varphi$ are expressed by the harmonic oscillator wave functions
 $R_{nLM}(\Vec{q},\beta_N)$ with the orbital angular momentum $L$, magnetic quantum number $M$,
 quanta $Q=2n+L$ $(n=0,1,\cdots)$, and size parameter $\beta_N=\sqrt{2N/(N-1)}\nu$.
The eigenvalues or occupation probabilities $\lambda$ are given as~\cite{gajda00}
\begin{eqnarray}
\lambda^{(LM)}_{n,N} = \frac{(4N)^{3/2}(N-2)^{2n+L}}{[3N-2+2\sqrt{2N(N-1)}]^{2n+L+3/2}},
\end{eqnarray}
and satisfy the following completeness relation,
\begin{eqnarray}
\sum_{L=0}^{\infty} \sum_{M=-L}^{L} \sum_{n=0}^{\infty} \lambda^{(LM)}_{n,N} =1.
\end{eqnarray}
The occupation probability with respect to the partial wave with quantum number $L$ is defined as
\begin{eqnarray}
\Lambda^{(L)}_N = \sum_{M=-L}^{L} \sum_{n=0}^{\infty} \lambda^{(LM)}_{n,N}.
\label{eq:PP_occupation_probablity_N_L}
\end{eqnarray}
In the macroscopic limit 
\begin{eqnarray}
\Lambda^{(L)}_{N=\infty} = (2L+1)(2-\sqrt{2})(3-2\sqrt{2})^L.
\label{eq:eq:one_body_den_int_infinite_probability}
\end{eqnarray}
We remark that the summed eigenvalues $\Lambda^{(L)}_{N=\infty}$ still depend
 on the angular momentum $L$ (see Fig.~\ref{fig:1}). 

\begin{center}
{\it Jacobi-type internal one-particle density matrix}
\end{center}

For the $N$-particle system, we define the $N-1$ internal Jacobi coordinates $\{ \Vec{\xi}_i \}_{i=1}^{N-1}$ and 
 the center-of-mass coordinate $\Vec{R}$ as follows:
\begin{eqnarray}
\Vec{\xi}_i = \Vec{r}_{N-i+1} -{\frac{1}{N-i}\sum_{k=1}^{N-i}\Vec{r}_k},\hspace*{3mm}\Vec{R} = {\frac{1}{N}\sum_{i=1}^{N} \Vec{r}_i},
\label{eq:Jacobi_coordinates}
\end{eqnarray}
where $\Vec{\xi}_1$ denotes the relative coordinate between the $N$-th particle and the remaining $(N-1)$ particles,
and other Jacobi coordinates are self-evident.
Then, the $N$-particle Hamiltonian in Eq.~(\ref{eq:total_hamiltonian}) can be separated into the internal and
 center-of-mass Hamiltonians, $H = H_{\rm int} + H_{\rm cm}$, where
\begin{eqnarray}
H_{\rm int}= {\sum_{i=1}^{N-1} \frac{\Vec{\pi}_i^2}{2\mu_i}}
       + {\sum_{i=1}^{N-1}\frac{\mu_i{\omega}^2{\Vec{\xi}_i^2}}{2}}, \hspace*{1mm} \mu_i=\frac{N-i}{N-i+1}m,
\label{eq:Jacobi_internal_Hamiltonian}
\end{eqnarray}
$\Vec{\pi}_i$ denotes the conjugate momenta with respect to coordinate $\Vec{\xi}_i$, 
 and $H_{\rm cm} $ is given in Eq.~(\ref{eq:Hamiltonian_cm}).
 
The Jacobi-type one-particle density matrix with respect to $\Vec{\xi}_1$ and $\Vec{\xi}_1^{\prime}$
 is given as~\cite{yamada08},
\begin{eqnarray}
\rho_{\rm int,J}^{(1)}(\Vec{\xi},\Vec{\xi}^\prime) = \int {d\Vec{\xi}_{2} \cdots d\Vec{\xi}_{N-1}}\hspace*{25mm} \nonumber \\
     \times~\Phi_{\rm int,J}^*(\Vec{\xi},\{\Vec{\xi}\}_{i=2}^{N-1})
            \Phi_{\rm int,J}(\Vec{\xi}^{\prime},\{{\Vec{\xi}}\}_{i=2}^{N-1}),\hspace*{2mm}\\
  = \left( \frac{N-1}{N} \frac{2\nu}{\pi} \right)^{3/2}
               \exp\left[-\frac{N-1}{N}\nu(\Vec{\xi}^2+{\Vec{\xi}^\prime}^2)\right],
\label{eq:Jacobi_one_body_den_int} 
\end{eqnarray}
where $\Phi_{\rm int,J}$ represents the internal, fully symmetric~\cite{matsumura04,yamada05,yamada08},
 wave function in Jacobi coordinates corresponding to Eq.~(\ref{eq:total_wf}).
This choice of the coordinate $\Vec{\xi}_1$ for the internal density matrix is natural, because
 the single particle orbit should be defined with respect to the relative coordinate between one particle and
 the other remaining $N-1$ particles in the Jacobi coordinate system. 
The separability with respect to $\Vec{\xi}$ and $\Vec{\xi}^{\prime}$ in Eq~(\ref{eq:Jacobi_one_body_den_int})
 stems from the fact that Jacobi coordinates form an orthogonal system 
 (it is noted that the cross term $\Vec{q}\cdot\Vec{q}^{\prime}$ does appear in the case of PP
 [see (\ref{eq:one_body_den_int})]
 because the internal coordinates used by PP are not orthogonal).
The eigenvalue equation of the one-particle density matrix can be solved analytically.
We find that the density matrix has only one eigenfunction $\varphi_{{\rm int,J},N}=R_{000}(\Vec{\xi},(N-1)\nu/N)$
 with non-zero eigenvalue, 
 namely, the $0S$ harmonic oscillator wave function with 100~\% occupancy, that is, in the same notation as above
\begin{eqnarray}
\Lambda^{(L)}_{N}=\sum_{M=-L}^{L} \lambda^{(LM)}=\delta_{L0}.\label{eq:Jacobi_occupation_prob}
\end{eqnarray}
This means that all particles are condensed in the single $0S$ particle state independent of $N$, 
 although the size parameter
 in the eigenstate $\varphi_{{\rm int,J},N}$ depends on $N$ and is slightly different from that in the
 eigenfunction $R_{000}(\Vec{r},\nu)$ in the laboratory frame, 
 as discussed above.

We, therefore, see that the eigenvalues and eigenfunctions of the internal 
density matrix depend on the choice of the internal coordinates. 
This is a somewhat surprising result.
In order to overcome the difficulty, we give a criterion for
 the choice of the internal coordinates: In the macroscopic limit ($N\rightarrow \infty$)
 the internal density matrix should have the same eigenvalues and eigenfunctions
 as those of the corresponding ideal Bose-Einstein condensate in the laboratory frame.
This is a very physical boundary condition.
Obviously, the PP-type one-body density matrix does not satisfy the condition, while
 the density matrix of the Jacobi-type does.
These results mean that one should take internal coordinates which do not produce any
 correlation in the internal one-particle density matrix in the macroscopic limit.
Otherwise, unphysical situations occur like the density matrix of the 
 PP-type which clearly is inadequate.

The present considerations can be applied to a general case of the internal one-particle density matrix
 for interacting bosons in a self-bound system.
The internal Hamiltonian of the system or translationally invariant shell-model
 Hamiltonian with the Jacobi coordinates (\ref{eq:Jacobi_coordinates})
 is presented as $H_{\rm int} = H - H_{\rm cm}$,
\begin{eqnarray}
H_{\rm int} = {\sum_{i=1}^{N-1} \frac{{\Vec{\pi}_i}^2}{2\mu_i}} + {\sum_{i=1}^{N-1} \frac{\mu_i\omega^2{\Vec{\xi}_i}^2}{2}}
        + \sum_{i<j=1}^{N} v(\Vec{r}_i-\Vec{r}_j),\hspace*{2mm}
\label{eq:internal_hamiltonian_interaction}
\end{eqnarray}
where the center-of-mass Hamiltonian $H_{\rm cm}$ of Eq.~(\ref{eq:Hamiltonian_cm})
 is subtracted, and $v$ presents the residual two-body interaction between bosons.
Defining $\Phi_{\rm int}(\{\Vec{\xi}_i\}_{i=1}^{N-1})$ as the eigenfunction
 of $H_{\rm int}$, the internal one-particle density matrix of the system is presented as\\[-6mm]
\begin{eqnarray}
\rho_{\rm int}^{(1)}(\Vec{\xi},\Vec{\xi}^\prime) = 
          \int \prod_{i=2}^{N-1} d\Vec{\xi}_i~ \hspace*{25mm}\nonumber \\
      \times~\Phi_{\rm int}^{*}(\Vec{\xi},\{\Vec{\xi}_i\}_{i=2}^{N-1})\Phi_{\rm int}(\Vec{\xi}^{\prime},\{\Vec{\xi}_i\}_{i=2}^{N-1}),
\label{eq:internal_one_particle_den_v_xi}
\end{eqnarray}
which is similar to Eq.~(\ref{eq:Jacobi_one_body_den_int}).
This definition satisfies the physical boundary condition in the limit of $N\rightarrow \infty$
 in the case of the absence of the interactions among bosons.
Full symmetry of the internal one-particle density matrix (\ref{eq:internal_one_particle_den_v_xi})
 can be verified~\cite{matsumura04,yamada05,suzuki02}.

In conclusion, we discussed the internal one-particle density matrix for Bose
 condensates in a harmonic trap and for self-bound Bose-systems.
It was found that the PP-type one-particle density matrix is
 physically inadequate to study the internal degree of Bose condensation, 
 while the density matrix of the Jacobi-type is fully appropriate. 
The use of the latter is, thus, of great importance for the exploration of
 condensates in self-bound
 systems such as $\alpha$-particle condensates in nuclei or small
 droplets of superfluid $^4$He, where only internal degrees of freedom
 are relevant.
Indeed, the definition (\ref{eq:internal_one_particle_den_v_xi}) has 
successfully
 been applied to study the degree of Bose condensation for self-bound $\alpha$-
particles in nuclei, unambiguously demonstrating
 that self-conjugate nuclei such as $^{12}$C, $^{16}$O, $\cdots$ 
show long-lived excited states close to the $n\alpha$ disintegration threshold 
where the $n \alpha$-particles form, in good approximation, a product state 
of 0$S$-orbitals, that is a condensate~\cite{matsumura04,yamada05,funaki08,
suzuki02,nupecc08}.


\end{document}